\RequirePackage{fix-cm}
\documentclass[smallextended]{svjour3}       
\smartqed  
\usepackage{graphicx}
\usepackage{mathptmx}      
\usepackage[hidelinks,breaklinks]{hyperref}
\usepackage[hyphenbreaks]{breakurl}

\usepackage{color}

\journalname{---}

\begin{document}

\title{Essay Review of Tanya and Jeffrey Bub's \emph{Totally Random: Why Nobody Understands Quantum Mechanics: A Serious Comic on Entanglement\thanks{Cuffaro gratefully acknowledges support from the \emph{Alexander von Humboldt Stiftung.}}}}

\titlerunning{Essay review of \emph{Totally Random}}        

\author{Michael E. Cuffaro         \and
        Emerson P. Doyle 
}


\institute{M. E. Cuffaro \at
  Munich Center for Mathematical Philosophy \\
  Geschwister-Scholl-Platz 1 \\
  D-80539 M\"unchen \\
  Tel.: + 49 (0) 89 / 2180 - 3319\\
  \email{mike@michaelcuffaro.com}           
  \and
  E. P. Doyle \at
  Huron University College\\
  1349 Western Rd.\\
  London, ON N6G 1H3\\
  \email{emerson.p.doyle@gmail.com}
}

\date{Preprint submitted to \emph{Foundations of Physics}}

\maketitle

\begin{abstract}
  \begin{sloppypar}
    This is an extended essay review of Tanya and Jeffrey Bub's \emph{Totally Random: Why Nobody Understands Quantum Mechanics: A serious comic on entanglement}. Princeton and Oxford: Princeton University Press (2018), ISBN: 9780691176956, 272 pp., 7$\times$10 in., 254 b/w illus., \textsterling 18.99 / \$22.95 (paperback). We review the philosophical aspects of the book, provide suggestions for instructors on how to use the book in a class setting, and evaluate the authors' artistic choices in the context of comics theory.
    \end{sloppypar}
\end{abstract}

\section{Introduction}

Tanya and Jeffrey Bub's \emph{Totally Random: Why Nobody Understands Quantum Mechanics} is a beautiful book, conceptually, artistically, and in the way that the concepts and art are combined to engage with the reader in a meaningful way. The reader is literally a character in this book, and as we will elaborate on in the course of this review, that is a large part of the book's point. We recommend it, in the highest terms, to instructors seeking to supplement an introductory course on quantum mechanics at the undergraduate or graduate level, to students in courses like this looking for a way to solidify what they have been learning with something fun and approachable, and perhaps most of all to those who are not even enrolled in a university course, but who have always wanted to (really) learn something about quantum mechanics. Even those who are already experts in quantum mechanics will find this to be an engaging and entertaining read. In short, we recommend this book to \emph{anyone} who is interested in learning about quantum mechanics, and serious enough in that interest to, with a paper and pencil in hand, work with the book's other characters through some elementary logic and arithmetic. While, as the book's subtitle indicates, nobody really understands quantum mechanics, anyone who reads this book will come away from it with a good understanding of just \emph{what} it is about quantum mechanics that is so hard to understand.\nocite{bell2004}

As Bub \& Bub literally illustrate, this is quantum entanglement, or as they put it in \emph{Totally Random}: It is the ``curious correlation'' between ``entangled quoins'' that they introduce the reader to in Part I, vigorously debate the foundational and philosophical significance of with cartoon-ish versions of the founders of quantum mechanics in Part II, and explain the practical applications of in Part III. And by focusing on quantum entanglement, Bub \& Bub are able to successfully get across to the reader of this book a lesson that Jeff Bub, in numerous of his publications, has been conveying to us for years: that what is new and surprising about quantum mechanics (in relation to the theories that it replaces) is the novel set of constraints that it imposes on our representations of phenomena. This is what Jeff Bub's `information-theoretic interpretation' of quantum mechanics amounts to, and although \emph{Totally Random} does not explicitly defend that view (or any particular view), it is, as we will explain further below, a beautiful expression and illustration of it and its value.

In the first section below we elaborate on these interpretive underpinnings, especially in comparison to Jeff Bub's previous work, \emph{Bananaworld: Quantum Mechanics for Primates} \cite{bub2016}. Section three turns to a brief discussion of the pedagogical aims and value of the book, along with parsing recommendations for instructors and lay readers alike. We argue that what makes \emph{Totally Random} so effective, both pedagogically and as downright entertaining reading, is a keen awareness and deployment of the strengths of its chosen medium. This point is more fully explored in the final section, where we employ comics theory to analyze the book \emph{as} a comic in its own right, and in comparison to other recent attempts to present difficult subjects in this atypical format. We conclude that its success and uniqueness among its peers can be attributed in large part to the authors' decision to include the reader as a genuine character in the book; a conceit for which comics as a medium are uniquely suited.

\section{Quoin mechanics according to Bubism}

A reader not already familiar with \emph{Jeffrey} Bub's previous work in the foundations and philosophy of quantum mechanics will probably not get the impression from reading \emph{Totally Random} that any particular interpretation of quantum mechanics is being advocated for in the book. This is appropriate for a book of its kind, aimed as it is at a general audience. It is true that the views of some of its characters, especially Bohr-(ish), are presented more favourably by the authors of \emph{Totally Random} than they are in many presentations (both popular and academic) of quantum mechanics. But ``more favourably'' here just means more fairly. Neither Bohr-ish's view nor any view is defended overtly in the book. As for the reader who \emph{is} already familiar with Jeff Bub's previous work, especially \emph{Bananaworld: Quantum Mechanics for Primates} \cite{bub2016}, it will be clear that the information-theoretic interpretation of quantum theory defended in \emph{Bananaworld} suffuses \emph{Totally Random}, in the way that Bub \& Bub introduce the strange kinematics of quoins in Part I, in the way that they frame the early debates about the interpretation of quantum mechanics in Part II, and finally in the way that they illustrate the practical use that can be made of nonlocal correlations in Part III. But it is in the mode of presentation of the book, not in its content, that Jeff Bub's information-theoretic interpretation makes itself felt.

\emph{Totally Random} is of course a comic book, not a (traditional) textbook, and not a philosophical treatise. But Bub \& Bub succeed in getting across to the motivated reader, who need only be familiar with the most basic elementary arithmetic to follow along, how (super-)quantum nonlocal correlations differ from classical correlations; how quantum puzzles that the reader will have heard of, such as ``Schr\"odinger's cat,'' ``spooky action at a distance,'' ``quantum teleportation,'' and so on, stem from these differences; and finally how practical use can be made of nonlocal correlations for computing and communicating information. And although no particular interpretation of quantum mechanics is defended overtly in the book, there is, nevertheless, an argument that can be made on the basis of \emph{Totally Random} for Jeff Bub's information-theoretic interpretation of quantum mechanics, and it is this: The book's very success in faithfully conveying to a completely general audience (in a comic book of all things!) the most puzzling aspects of quantum theory, highlights what Jeff Bub has been telling us for years: that what is novel about quantum mechanics' descriptions of phenomena (as compared with the descriptions of phenomena given by classical mechanics) is the framework of logico-probabilistic constraints that are imposed on them by the theory. That, in fact, is what the label `information-theoretic' means. Quantum mechanics is \emph{about information} in the sense that what is \emph{novel} about the theory can be expressed in information-theoretic terms. This suggests that we think about the practical use that can be made of quantum-mechanical systems, and as Bub \& Bub explain in Part III, it provides us with guidance for how to do so (see also \cite{bub2006,bub2008,bub2010}). It also allows the very successes achieved in these practical applications to inform our understanding of the theory's foundations (see, e.g., \cite{bub2012}), and in general invites a mutually fruitful interchange of ideas between philosophy and work at the cutting edge of the foundations of quantum mechanics.

This focus on comparing the logico-probabilistic constraints imposed on descriptions of phenomena by quantum and classical mechanics (where the latter is represented abstractly by the more general concept of a local hidden-variable theory) forms the core of the approach pioneered by John Bell \cite{bell1964,bell1966,bell1981}. For Bell, however, this approach to interpreting quantum mechanics is no more than an (ingenious) starting point for a more traditional deeper investigation into the ontology of quantum systems \cite{bell1984,bell1992}. Itamar Pitowsky, in contrast, discerns something important in this approach in its own right, follows it through to its logico-probabilistic conclusions \cite{pitowsky1989}, and draws a different moral for our understanding of the world related to us by quantum mechanics than the one drawn by Bell; a moral which Jeff Bub, foremost among a number of authors, has since continued to develop and interpret in the light of advances in fundamental quantum physics,\footnote{Others who advocate for a similar approach are William Demopoulos (especially in the book \emph{On Theories} \cite{demopoulosForth} that he completed shortly before his death in 2017) as well as Michael Janas, Michael Cuffaro, and Michel Janssen \cite{3m2020}, who go so far as to dub the interpretation of quantum mechanics defended in their own book as a version of ``Bubism'' (not to be confused with QBism), a label
originally coined by Robert Rynasiewicz (private communication).} advances that if anything seem more and more to confirm Bub \& Pitowsky's basic insight that the novel information-theoretic content of quantum mechanics is the key to the theory's deeper significance for our understanding of the world.

\begin{sloppypar}
Quantum mechanics imposes a particular set of physically motivated logico-probabilistic constraints, implied by the Hilbert space structure of the theory, on how the observable quantities associated with physical systems relate to one another. Famously, correlations between physical systems, in light of these constraints, are in general nonlocal, a fact that has been and continues to be confirmed experimentally. There is something else very special (even if not unique) about the kind of nonlocal correlations that quantum mechanics allows, though: Quantum-mechanical correlations, despite being nonlocal, are also provably \emph{non-signalling}. This means that the (marginal) probabilities that Alice should assign to the possible results of her local experiments on a subsystem of a given system do not depend on anything Bob does or does not do to another subsystem of the system spatially separated from hers. This is true even when Alice's and Bob's subsystems are entangled with one another.
\end{sloppypar}

As special and physically significant as the non-signalling condition is, however, quantum mechanics is not alone among conceivable physical theories in enforcing it. The non-signalling condition also provably constrains, in particular, theories in which `PR-boxes' \cite{popescu1994} are possible. The nonlocal correlations exhibited by PR-boxes are actually stronger than those allowed in quantum mechanics. They are, as it turns out, the strongest possible correlations that can be conceived of that still satisfy no-signalling. What does this have to do with \emph{Totally Random}? As Bub \& Bub point out on p. 8, the ``curious correlation'' that is the actual subject matter of their graphical experiential narrative is not the one exhibited by quantum-mechanical systems. It is the one exhibited by PR-boxes, which take the form, in Bub \& Bub's narrative, of `superquantum entangled quoins', the output of the \emph{Super Quantum Entangler PR01}, which as ``J'' (Jeff Bub's superego alter ego) points out, on p. 8, looks a lot like a toaster.

Even though quoin correlations are not quantum correlations, they share with quantum correlations the main feature of the latter that gives rise, from the classical point of view, to the familiar puzzles and paradoxes of quantum theory. As ``T'' (Tanya Bub's id) progressively discovers as she makes her way through Part I of the narrative, these seemingly banal quoins manifest correlations that defy her attempts to explain them. They are not explainable in terms of a common cause, unless our own choices about which way we flip them are already determined in advance of our actually doing so, nor do they directly influence one another, on pain of contradiction with everything we think we know about physics. The only alternative seems to be that, ``while there are rules that describe the correlations, there is no cause that makes them happen. They just sort of happen to happen, by chance, in a coincidentalish, totally random sort of a way'' (pp. 64---65).

There is more to be said, and in the remainder of the book Bub \& Bub expand upon the conundrum presented by quoin mechanics. But before moving on we should comment on quoins. If the novel content of quantum mechanics can, as we mentioned, be expressed in information-theoretic terms, what is gained by focusing on super-quantum as opposed to quantum correlations? Indeed, it is certainly possible to write a book introducing quantum mechanics to the physically and mathematically uninitiated,\footnote{For instance, see \cite{jordan2005}. Parts of \cite{3m2020} (especially ch. 2) are also written in this way.} but it is hard to imagine doing so (especially in a comic book) if the reader one has in mind has little to no prior understanding of the basic mathematics of probability theory (or has long forgotten it). This is because quantum-mechanical correlations violate the assumptions of local hidden-variable theories in an essentially probabilistic way. It is true that the original version of Albert Einstein, Boris Podolsky, and Nathan Rosen (EPR)'s \emph{Gedankenexperiment} \cite{epr1935} was cast in terms of deterministic correlations. But thanks to Bell \cite{bell1964} we now know that the particular fragment of the correlational space of EPR's setup is recoverable in a local hidden-variable theory, and that it is only when we move to the \emph{statistics} produced by an EPR pair, for \emph{other} measurement settings besides the ones given in the EPR paper, that a contradiction between those statistics and the assumptions of local hidden-variable theories emerges. Even then, such a contradiction only emerges with increasing confidence, not with certainty, as the number of repetitions of the experiment increases. True, in the Greenberger-Horne-Zeilinger (GHZ) setup \cite{ghz1989,ghsz1990}, unlike in Bell's version of the EPR setup, the departure from the assumptions of local hidden-variable theories is deterministic, not probabilistic. But the GHZ setup introduces the extra complication of a third system, among other things.

As for quoins, they produce deterministic correlations that are provably both non-signalling and incompatible with the assumptions of any local hidden-variable theory despite the fact that they only involve two separated systems. The correlations between two such systems can, moreover, be completely described in an extremely concise way (see the visual on p. 15 of \emph{Totally Random})---and the fact that they violate the assumptions underlying local hidden-variable theories can very easily be proven (p. 41). Further, as the point of the book is not to distinguish quantum from super-quantum theories, but to convey the basic gist of the way that quantum correlations depart from classical correlations---a difference with classical mechanics that both quantum and super-quantum theories have in common (though to different degrees)---not much is lost, conceptually, by appealing to quoins to elucidate this difference. Finally, we note that probabilities are not entirely absent from the book, after all. They enter, albeit in a mostly trivial way, via the fact that quoin correlations satisfy the non-signalling condition. Thus the marginal probability for one quoin to land heads, irrespective of what is done with the other quoin, is 50\%. Presumably, verifying that this is indeed the case is what the many repetitions of the quoin experiment are for on pages 19 and 55.

Popescu-Rohrlich correlations play a large role in \emph{Bananaworld}, just as they do in \emph{Totally Random}. But in \emph{Bananaworld} they are modelled, not by quoins, but (as the title of the earlier book indicates) by Popescu-Rohrlich \emph{bananas}. This invites the question: Why the change to quoins? As Sandu Popescu indicates in the foreword to the earlier book, bananas certainly have one advantage:

\begin{quote}
The more I think of the bananas, the more I like their use: The two peeling actions are complementary and cannot both be implemented on the same banana---once it's peeled it's peeled, exactly like two measurements that cannot be simultaneously performed on microscopic particles. And there is only one answer---once it's eaten it's eaten. Again, exactly like in quantum mechanics. The magic bananas are a perfect model for what is going on. So much better than the abstract models the physicists use \cite[vi]{popescu2012}.
\end{quote}

We also like bananas for the same reason. But one cannot say the same thing of quoins. Of course one can stipulate that quoins only work on the first flip, but this is artificial; a quoin looks and feels the same way both before and after we flip it. That said, this fact about quantum mechanics (and bananas) is not really needed for the purposes of the exposition in \emph{Totally Random}. For the purposes of that exposition a quoin works just as well. Moreover a quoin conveys the idea of a random process much better than a banana does; and in general it is not clear how bananas could have been made to fit in with the narrative of \emph{Totally Random}. It is hard to see, in particular, how bananas could have been made to work as well as quoins in the casino scenes in Part III (or, for that matter, in Part I),\footnote{\emph{Bananaworld} does contain a few descriptions and depictions of games with bananas. While we love the illustrations (see, e.g., pp. 195--196), quoins are clearly more natural in the context of such games than bananas are.} though perhaps one can just as easily imagine Einstein-ish, Bohr-ish, and friends obsessing over bananas instead of quoins in Part II.

Regarding these various ``-ishes,'' it is worth noting that only the real-life historical counterparts of Einstein-ish and Schr\"odinger-ish really engaged with the phenomenon of quantum entanglement in their writings on quantum mechanics. As we noted above, Einstein, with Podolsky and Rosen, first drew physicists' attention to the phenomenon in their paper of 1935, while Erwin Schr\"odinger, in his three-part commentary on the EPR paper from the same year, actually coined the term `entanglement', presenting it as, not one but ``\emph{the} characteristic trait of quantum mechanics, the one that enforces its entire departure from classical lines of thought'' \cite[555, quoted and translated in \emph{Totally Random}, p. 9]{schrodinger1935}.\footnote{The phenomenon is also discussed by Grete Hermann \cite[276]{hermann1935a}, though she does not actually use the word `entanglement' in her own commentary since it was Schr\"odinger who coined the term in his paper from the same year.} After Bohr's reply to EPR \cite{bohr1935}, however, most physicists of the day (who professed to be convinced by it) came to the conclusion that the puzzling features of quantum entanglement were merely an artefact of Schr\"odinger's and Einstein's outmoded views of physics. It was only after the work of Bell, some thirty years later, that physicists and philosophers came to be interested in quantum entanglement again. It therefore distorts the actual history somewhat to present the other ``-ishes'' who appear in Part II as preoccupied with the curious correlations of quoin mechanics, as opposed to the other puzzling aspects of the new theory. But as we previously noted, the upshot of \emph{Totally Random} is, to paraphrase Schr\"odinger, that the curious correlation is behind all of these puzzles, even if, for the Bubist, Einstein's and Schr\"odinger's \emph{rejections} of quantum mechanics did indeed stem from their mistaken \emph{a priori} conceptions of what a fundamental physical theory should be like.

Quoin mechanics (and quantum mechanics) are fundamentally different from classical mechanics. How so? The Bubist story goes something like this: Imagine all of the possible Boolean (i.e., yes-or-no) questions that can be asked about a particular observable quantity, $A$, associated with a system at a particular time, questions like: \emph{Is the value of the observable quantity $A$ within the range $\Delta$?}. The classical state description constitutes a \emph{truthmaker} \cite[433]{bub-pitowsky2010} in relation to that observable in the sense that, once a state assignment is made, the answers to all such questions about the quantity---which together can be represented as a Boolean algebra---are determined in advance. Further, in classical mechanics, the Boolean algebras corresponding to all of the individual observables associated with a system can be embedded into a larger globally Boolean algebra comprising them all. Thus, in classical mechanics, once we assign a state to a system, every yes-or-no question concerning \emph{any} of its observable properties is determined in advance, irrespective of whether we ask that question or not, and irrespective of whatever other questions we happen to ask about the system. We are thus invited to think of the values of these observable quantities as \emph{properties} of the system that it possesses independently of anything we may or may not do to it.

Likewise for questions concerning combinations of observable quantities, questions like: \emph{Is the value of the observable quantity $A\times B$ within the range $\Delta$?} The answer to such a question is completely determined given the individual values of $A$ and $B$, which are in turn completely determined once the system is assigned a state. The same goes for a system composed of many parts, for instance a bipartite system $C$ composed of two American coins $C_1$ and $C_2$, since any observable property $A$ of the subsystem $C_1$ is also an observable property of $C$, and similarly for any observable property $B$ of $C_2$. If we correlate $C_1$ and $C_2$ using Schr\"odinger-ish's \emph{Correlator Classique} \cite[89]{bub2018}, and then ask his \emph{So-Sein} machine: \emph{Has $C_1$ been rigged by the Correlator Classique to land tails?}, then given the \emph{So-Sein} machine's answer one can, using \emph{coin mechanics} (p. 88), determine what the outcome of tossing both coins will be even before actually tossing them. The properties possessed by $C_1$, which the \emph{So-Sein} machine has helpfully determined for us, combined with what we know about the coin mechanics of the \emph{Correlator Classique}, are such that it has to be so.

How about quoin mechanics? States in quoin mechanics fail to be truthmakers in relation to the observable quantities associated with a system in two senses. First, assigning a state to a system yields, in general, only a particular probability that a given observable will take on a particular value when we query the system concerning it. Consider a system $Q$ composed of two entangled quoins, $Q_1$ and $Q_2$, output from the \emph{Super Quantum Entangler PR01}. Let $H_1$ represent the observable quantity associated with a toss of $Q_1$ that begins with heads facing up, $T_1$ be the observable quantity associated with a toss of $Q_1$ that begins with tails facing up, and let $H_2$ and $T_2$ be the corresponding observables for $Q_2$. The state of $Q$ fails to be a truthmaker with respect to the values of the observable $H_1$, for instance, in the sense that the outcome of the corresponding toss will be totally random, and likewise for $T_1$, $H_2$, and $T_2$. Now, we \emph{can} associate a classical (50/50) probability distribution with each of these observables, and in that sense we can associate a Boolean algebra of properties with each of $H_1$, $H_2$, $T_1$, and $T_2$, even if those properties fail to determine the outcome of the corresponding toss with certainty.

This brings us to the second, more important, sense in which the state of a pair of quoins, $Q$, in quoin mechanics, fails to be a truthmaker with respect to the observables associated with $Q$: In quoin mechanics, the Boolean algebras corresponding to individual observables cannot be thought of as sub-algebras of a globally Boolean algebra; i.e., there is no way to assign Boolean algebras to the individual observables $H_1$, $H_2$, $T_1$, $T_2$, $H_1T_1$, $H_1T_2$, $\dots$ consistently so that they all fit together into one globally Boolean algebra (\cite[41]{bub2018}; see also \cite{abramsky2015}). Rather, a Boolean algebra is assigned to an individual observable \emph{conditional upon the selection} of that observable. In other words quoin mechanics \emph{does not} associate a Boolean algebra to a given observable \emph{in advance} of a question concerning that observable having been asked. According to quoin mechanics, if we want to get a meaningful answer from nature we actually have to ask it a question.

This is the picture that Einstein-ish, in his characteristic way, rejects. For Einstein-ish it is inconceivable that physics should describe the world in this manner. For Eintein-ish physics just \emph{is} the enterprise to ascribe an independent reality or ``being-thus''---in other words a Boolean algebra of properties---responsible for the values obtained in a quoin flip. And given this assumption, quoin mechanics simply must be rejected because it implies, for two quoins $A$ and $B$, ``that a physical reality in `$B$' undergoes an instantaneous change because of an action made on `$A$'.'' Einstein-ish refuses to accept this.

Among the characters that make an appearance in Part II, the one whose view most closely represents Jeff Bub's own is Bohr-ish \cite{bub2017}. Bohr-ish exhorts us to let go of the conception of reality as ``a given thing, all of whose aspects can be viewed or articulated at any given moment'' (p. 155) that leads Einstein-ish and Schr\"odinger-ish to reject quoin mechanics, and that leads Everett-ish to the conclusion that what seems to us like a single world is in reality many (p. 154). Bub \& Bub's depiction of Bohr-ish as a psychoanalyst, however, conveys to the reader that it is not easy to give up on Booleanity. What could it possibly mean to live in a non-Boolean world? The idea that whatever underlies our experience can be conceived of as something independently existing runs deep. It underlies our common understanding of what it means to explain in physics.

Bohr-ish nevertheless enjoins us to move beyond it. The world according to quoin mechanics, as conceived by Bohr-ish (and by the Bubist), is \emph{not} the conception rejected as paradoxical by Schr\"odinger-ish---a world in which comic book heroes can be both dead and alive at once, and in which ``an object [can be] rigged to land heads \emph{and} rigged to land tails \emph{at the same time''}---the conception ultimately embraced by Everett-ish through the clever manoeuvre of multiplying the world into just enough copies to resolve all of the paradoxes that Schr\"odinger-ish cannot cope with. The world according to quoin mechanics as conceived by the Bubist is a world in which our experimental choices actually matter, and have a bearing on the future experimental possibilities that are open to us. In the world as conceived by the Bubist there is only one story to tell, not many, as Everett-ish would have it. But it is a ``choose your own adventure'' story like the one related by T. on pp. 98--101, a story that we can perfectly well conceive of and use physics to describe and investigate in the way that we always have, but in which the choices that we make cannot all be put together consistently into one overarching picture of (an underlying) reality. As T. puts it, ``there's simply no sort of hero that can make all the endings true at the same time!'' (p. 101).

\section{Suitability to Instructors and General Audiences}

As we mentioned at the outset, \emph{Totally Random} is an easy recommendation for instructors wanting to supplement an introduction to quantum mechanics or philosophy of quantum mechanics course at the undergraduate or graduate level with something less technical and more fun or approachable. While the book is unlikely to be thorough or deep enough to act as the sole text for such a course, the narrative is mostly structured so as to make for easy assignment of discrete parts on a regular schedule alongside more rigorous presentations of those same topics.

As an example of this structure we may return to the ``-ishes'' of Part II. Each is a delightful caricature---`Schr\"odinger-ish'' as a buttoned-up but deranged engineer with far too many cats, ``Everett-ish'' as a slick carnival barker, and so on---who introduces their favoured interpretation by way of attempting to help the reader and ever-present narrator understand quoin mechanics. These scenes are playful, engaging, and informative; but always before too long we are whisked away by some narrative device to the next scene, and so next interpretation. This culminates at the end of Part II in a lively debate among all the caricatures within the pseudo-psychiatric offices of Dr. Bohr-ish, wherein Bohm-ish takes refuge in a closet from a very boisterous Einstein-ish. Bohr-ish manages the final word on interpretations---the authorial motivations for ending with this scene having been discussed above. Pedagogically the effects of this structure are discrete vignettes that act as excellent introductions and jumping-off points for further classroom discussion of each interpretation. The presentations of several of the most engaging potential applications in Part III are even more discrete yet narratively inventive.
 	
This quick, varied, and amusing structure means the book is also well-suited for self-study by a general audience. As mentioned above, the book is approachable enough that anyone with an interest in quantum mechanics will find this a solid first presentation. Of course, it's still a book on quantum mechanics, and so requires sustained effort and careful reading to gain the most from its pages.  This is especially the case for Part I. Although the device of quoin mechanics makes for an especially clear illustration of non-classical correlations,\footnote{See especially pp. 38--41 for an extremely elegant and vivid combinatorial presentation of the impossibility of replicating such behaviour classically---the frustration expressed by the narrator is palpable!} the part is quite long and has few natural stopping points. Still, even this longer part remains engaging, mystifying, and entertaining throughout, as the authors use their chosen medium quite aptly to encourage the reader to press on to the later parts, so that they may uncover some explanation for quoin mechanics and resolve the story.
	
A final very welcome addition for instructors and lay readers alike are the frequent double-page spreads which present facsimiles of prominent newspaper stories featuring quantum mechanics from throughout the 20th century, or show the first pages of key articles and correspondences from the history of quantum mechanics.  The authors' own self-injected caricatures annotate these spreads with glib conversation that works well to highlight the import of these materials or express solidarity with the reader that the details can be intellectually challenging but rewarding. Pedagogically these interjections serve not only to provide the reader even more natural pausing or reflection points, but also demonstrate to the uninitiated that these topics were of considerable newsworthiness in their time. Along similar lines, there are 10 full pages of endnotes which provide definitions, explanations, and point to key sources---nearly every scene in the book is thus annotated, with further expansion at the book's website. These notes are unfortunately tucked away in the back without so much as a reference. We almost missed them completely on our first reading. There is a wealth of supplementary information here, so leaving out any indication to this trove within the main work seems an unfortunate oversight.

That the book ultimately serve as an apt and unique pedagogical resource was clearly a driving force for the authors. In an interview available on the book's companion/promotional website (which also includes a wealth of introductory-level educational resources), Tanya Bub reinforces this idea: \begin{quote}What we wanted was for readers to have that ``Aha!'' moment of understanding when you experience something directly. Wouldn't it be cool if instead of just telling you about how weird quantum mechanics is, we could somehow hand you an object that has all the weirdness of quantum entanglement baked into it, so that you get to play with it and see for yourself. [\dots] That's when we came up with the idea of crafting a quantum object and making it ``real'' in the form of an experiential comic. \cite{web2020}\end{quote} Of special note is of course the choice of medium, which apparently developed organically from the short comic-panel sections developed for \emph{Bananaworld}. This choice is in an important way essential to the work, and so this review would be incomplete if we did not include an evaluation of \emph{Totally Random} qua comic book.

\section{Success as a ``Graphical Experiential Narrative''}

Whenever presenting material in an atypical medium, the operative question is: \emph{Why bother?} Prototypical examples are Plato's dialogues, and so the medium of the academic dialogue more generally. Leaving to one side historical circumstance regarding the scholarly and literary norms of ancient Greece, the medium of the dialogue does more than simply provide an entertaining forum for Plato to explore or explain his ideas. This choice of medium directly illustrates the Socratic method---teaching readers to do philosophy by the very structure of the work. In this sense the medium becomes an essential aspect of the work, and so becomes essential to the work's goals; the point being that the effectiveness of Plato's works as a tool of pedagogy would be greatly lessened were they presented in another medium---as standard philosophical prose, for example. \emph{Totally Random} achieves similar levels of integration with its chosen medium, taking advantage of both the dialogue-driven and illustrative potential of the comic book over a work of pure prose to a high degree.  The work thus distinguishes itself from the few noteworthy works which share its atypical mode of presentation.

The two obvious and immediate comparisons are the prolific \emph{Introducing\dots\ A Graphic Guide} series of graphic nonfiction books, and the immensely successful \emph{Logicomix} by Doxiadis and Papadimitriou, art by Papadatos and Di Donna \cite{doxiadix2009}. As with \emph{Totally Random}, these works aim to present conceptually difficult material to the lay or introductory reader by leveraging the more inviting and entertaining medium of the comic book. While all are successful in their own way, \emph{Totally Random} relies upon the special strengths of the comic book to present a ``quantum object'', as Tanya Bub says, to the reader and achieve its pedagogical goals more fully.

This point can be made more robust with a deeper comparison of \emph{Totally Random} to the other comic-type titles just mentioned, although this requires a short foray into the world of comic theory. In \emph{Understanding Comics: The Invisible Art} \cite{mccloud1994}, widely recognized as one of the paramount works of comic history and theory, author Scott McCloud tentatively defines comics as: \begin{quote}Juxtaposed pictorial and other images in deliberate sequence, intended to convey information and/or to produce an aesthetic response in the viewer. (p. 9) \end{quote} Deliberately broad, this definition includes a wide variety of works, from the ever-present comedy panels in the Sunday newspaper, to modern superhero comics, and all the way back to ancient cave paintings. Casting this widely allows us to better contextualize comic books in history and as a form of art. While all the titles mentioned in the previous paragraph meet McCloud's definition, that they do so in quite different ways is instructive. In chapter 6 of his study, McCloud pulls back from the comic book specifically to speak about art more generally, discussing the origins of art and its role as an aspect and expression of the human condition. He argues that all artists, regardless of medium, must make a choice about a given piece: ``does the artist want to say something about life \emph{through} his art or does he want to say something about \emph{art itself}?'' (p. 178). This choice reflects a dichotomy in the author's primary impulse or artistic goal for a piece---to focus on expressing an \emph{Idea}, or to push the boundaries of a medium's \emph{Form} itself. Creators lie on a spectrum of course, and the choice may sometimes be an unconscious one, but this dichotomy acts as a useful first-pass analysis and categorization of works in many mediums.

Examples of comic creators who are best categorized as \emph{Form-First} include French artist Jean ``Moebius'' Giraud,\footnote{Giraud is best-known for revolutionizing the American Western comic genre in the mid-60s with \emph{Blueberry}, as well as acting as storyboard developer for such films as \emph{Alien}, \emph{Tron}, and \emph{The Fifth Element}.} and the more experimental works of American artist Art Spiegelman. McCloud notes artists from other mediums that fall into this category include Igor Stravinsky, Virginia Woolf, and Orson Welles. Comics which more often deploy the typical conventions of the medium to better-convey an \emph{Idea}---some message or story---include Charles Schulz's \emph{Peanuts} strips and Belgian cartoonist Georges ``Herg\'e'' Prosper Remi's \emph{Tintin}. In other mediums McCloud references creators such as Charles Dickens, Woody Guthrie, and Edward R. Murrow as expressing \emph{Idea-First} sensibilities in their work.

To return now to the works of our comparison-class mentioned above, while all three exemplify an \emph{Idea-First} approach, \emph{Totally Random} does something much more substantial by tying its pedagogical purposes into the medium to such an extent. Contrast the long-running and somewhat storied \emph{Introducing\dots\ A Graphic Guide} series. This series has a significant pedigree, beginning with two Spanish-language entries \emph{Cuba para Principiantes} \cite{Rius1971} and \emph{Marx para Principiantes} \cite{Rius1979} by pioneering Mexican political cartoonist Eduardo ``Rius'' Humberto del Rio Garcia.\footnote{The first book was originally self-published by Rius in 1960 and has quite a storied publication history of its own. We have here cited a currently available English-language translation by Robert Pearlman. The second book was originally published in 1972, cited here is the very popular English-language translation by Richard Appignanesi which initiated development of the wider series in the United States.} At the time Rius' work was revolutionary---educational introductions to complex and controversial subjects presented in a radical, pop-art style. The second book especially features a mixture of provocative but cartoonish line-art alongside photographic cut-outs arranged in collage. Considerations of Form are exemplified by the works---pushing the boundaries of the comic medium at the time into the realm of the educational, the nonfiction. But as the series developed in the hands of other creators and editors, broadened its range of subjects, and other educational publishers ``caught up'' to the art style and presentation of such topics, these radical edges were dulled. So while modern entries remain entertaining and surprisingly comprehensive introductions to a great variety of topics,\footnote{Of particular interest to readers may be: \emph{Introducing Quantum Theory: A Graphic Guide} by J.P. McEvoy and Oscar Zarate \cite{Mcevoy2007} , \emph{Introducing Time: A Graphic Guide} by Craig Callender and Ralph Edney \cite{Callender2010}, \emph{Introducing Philosophy of Science: A Graphic Guide} by Ziauddin Sardar and Borin Van Loon \cite{Sardar2011}, and the entry with which the present authors are most familiar, \emph{Introducing Logic: A Graphic Guide} by Dan Cryan, Sharron Shatil, and Bill Mayblin \cite{Cryan2008}.} our key observation is that while the medium is often used in an illustrative and elucidatory way, there is nothing \emph{essential} to the works regarding the use of illustrations. Entertaining and engaging certainly, and often extremely helpful to illustrate a concept like set intersection, or to present a thinker's ideas directly from their own mouths in the form of speech-bubbles, but ultimately the effects are comparable to a graphically-inclined instructor's use of a PowerPoint presentation in class today. Considering this series overall, the emphasis on \emph{Form} that marks Rius' original works has gradually fallen away, partially as a function of standardization, and partially as a simple function of the artistic progress of other works.

Compare again \emph{Logicomix}, which is ultimately an historical novel.\footnote{An equally apt comparison is Spiegelman's \emph{Maus} \cite{Spiegelman1996} (serialized 1980--1991, and thereafter republished in single graphic novel format---the first graphic novel to win a Pulitzer Prize), which McCloud distinguishes as quite a departure from Spiegelman's other work in terms of style and focus, landing squarely on the \emph{Idea} side of the dichotomy in comparison to his more experimental works. Given the subject matter---an autobiography of the child of holocaust survivors---this more direct and ``report-style'' presentation conveys the lived experience of the author in a more deliberate fashion.}  The book follows the early life and career of Bertrand Russell as he struggles to develop and defend his logicist programme as a coherent foundation for mathematics, while also witnessing and engaging with the dramatic world events of the early 20th century. Papadatos and Di Donna's subtle and stylized art works to draw the reader into this drama and bring Russell's struggles and triumphs to life in a way that would be difficult with the written word alone. But as successful as the art in \emph{Logicomix} may be in this respect, again there is nothing about the story told which \emph{requires} it be presented as a graphic novel. Art arranged in panels make the story that much more vivid and engaging, but the narrative could be presented as well in another medium. So this is a book written squarely with an \emph{Idea} at the fore, ultimately using the form of the comic medium in a very traditional way. 

\emph{Totally Random} takes a decidedly different approach with the choice to introduce the reader as an active (but silent) participant, and this choice is inspired for a work meant to educate as well as entertain. The book is filled with glib remarks and tongue-in-cheek humor, but the first genuine smile prompted by the book comes right on the first page---in the ``Dramatis Personae'' section introducing the reader to the colourful cast with which we will engage. You, the reader, are introduced as a main character. Right there, fourth from the top, unsure about how this will play out in the rest of the book---what form your representation in the story will take---until you move your hands to turn the page, uncovering an illustrated pair of thumbs right where yours had been. Truly delightful to be sure, but more importantly it speaks to a narrative gambit Bub \& Bub have made in so integrating the book into its medium. For the most part the book takes the form of a \emph{conversation} that plays out not only among the sometimes substantial cast of characters on the page, but with the reader as active interlocutor. The authors have thus had to anticipate not just the \emph{content} of the reader's question or confusion at any given point in the text, but also anticipate the \emph{structure} that this question or confusion will possess in order not to confuse the conversational thread of the text. There are several cases wherein the characters are responding to something the reader has presumably asked or said, or where characters directly ask the reader to perform some action. Besides a pair of disembodied hands (or thumbs), the format requires leaving the lacuna in the conversational flow entirely as a matter of implicature. In almost every case this works very well, and so the reader partakes in an experience that would be fundamentally lacking from a work of pure prose.

These choices in presentation result in \emph{Totally Random} taking something of a middle-ground between a \emph{Form-} and \emph{Idea-First} approach, which we think helps to explain the book's pedagogical successes. A ``graphic experiential narrative'' seems uniquely suited to illustratively explore a complex topic while engaging in a dialogue \emph{with} the reader. Much like the \emph{Introducing\dots A Graphic Guide} series, the art in \emph{Totally Random} is (now digitally-composed) collage, with foreground elements and characters integrating pieces of heavily stylized photographs into line art, while backgrounds tend to be made up of simple textured patterns which alternately convey a minimal sense of place or---where necessary---a distinct lack of one. Characters, objects, and effects regularly violate panel boundaries, with this choice often being used as way to demonstrate confusion or a character becoming ``unhinged''.  When the Reader's hands or actions are represented on the page, they rightly completely ignore whatever panel structure a page might otherwise observe. 

Visual aids are well-used throughout to illustrate ideas that would otherwise require lengthy explanations. However there are a few places in the text where the illustrations---or more commonly, the position and flow of text that violates typical panel structure---is liable to confuse. Both in the discussion of the ``Quantum Casino'' and of cryptography in Part III, we found the flow of the dialogue between characters, narrator, and reader can be easily confused. It's not always entirely clear on a first pass who is asserting what. So this is one place where Bub \& Bub's gambit of anticipating the reader falls a bit flat. Similarly, the visual complexity of the illustrations depicting the aforementioned casino makes understanding the preparation for the odds-fixing game seemingly more difficult than it really is. On the other hand, we must remember that this is a book on quantum mechanics, and so the need to re-read a section or two is hardly a serious criticism of the book. That the remainder is so understandable, approachable, and entertaining is a testament to the idea that other philosophical and technical subjects could benefit greatly from similar treatments in this medium. Those undertaking such a task will find no better than Bub \& Bub's example.

\begin{acknowledgements}
  To be inserted
\end{acknowledgements}

%
%

\bibliographystyle{spmpsci}      
\bibliography{totally_random_review}   


\end{document}